\documentclass[final,3p,times]{elsarticle}

\usepackage{amssymb}

\usepackage{amssymb}
\usepackage[]{amsmath}
\usepackage{graphics}
\usepackage{amssymb}
\usepackage[]{amsmath}
\usepackage{amsthm}
\usepackage{float}
\usepackage{setspace}
\usepackage{epsfig}
\usepackage{subfigure}
\usepackage{booktabs}
\biboptions{numbers,sort&compress}
\usepackage{graphicx}
\makeatletter

\newcommand{\Rmnum}[1]{\expandafter\@slowromancap\romannumeral #1@}
\makeatother

\usepackage[dvipdfm,colorlinks,linkcolor=red,anchorcolor=blue,citecolor=green]{hyperref}

\usepackage{times}
\usepackage{anysize}
\marginsize{2.5cm}{2.5cm}{1cm}{1cm}

\selectfont   

\journal {}

\begin{document}

\begin{frontmatter}



\title{Localized nonlinear waves of the three-component coupled Hirota equation by the generalized Darboux transformation}

\author{Tao Xu}

\author{Yong Chen\corref{cor1}}
\ead{ychen@sei.ecnu.edu.cn}

\cortext[cor1]{Corresponding author.}

\address{\rm Shanghai Key Laboratory of Trustworthy Computing, East China Normal University, Shanghai, 200062, People's Republic of China}

\begin{abstract}
In this paper, We extend the two-component coupled Hirota equation to the three-component one, and reconstruct the Lax pair with $4\times4$ matrixes of this three-component coupled system including higher-order effects such as third-order dispersion, self-steepening and delayed nonlinear response. Combining the generalized Darboux transformation and a specific vector solution of this $4\times4$ matrix spectral problem, we study higher-order localized nonlinear waves in this three-component coupled system. Then, the semi-rational and multi-parametric solutions of this system are derived in our paper. Owing to these more free parameters in the interactional solutions than those in single- and two-component Hirota equation, this three-component coupled system has more abundant and fascinating localized nonlinear wave solutions structures. Besides, in the first- and second-order localized waves, we get a variety of new and appealing combinations among these three components $q_1, q_2$ and $q_3$. Instead of considering various arrangements of the three potential functions, we consider the same combination as the same type solution. Moreover, the phenomenon that these nonlinear localized waves merge with each other observably, may appears by increasing the absolute values of two free parameters $\alpha, \beta$. These results further uncover some striking dynamic structures in multi-component coupled system.

\end{abstract}

\end{frontmatter}

\section{Introduction}
In recent years, the semi-rational localized waves, which include bright or dark solitions, breathers and rogue waves, have been one of the fascinating topics which have some potential applications in Bose-Einstein condenstates in atomic physics, optical fibers in nonlinear optics and other fields. Rogue waves\cite{01,02,03,04}, (also called freak waves, monster waves, killer waves, rabid-dog waves) have peak amplitude usually more than twice the significant wave height, and also appear from nowhere and disappear without a trace. Breathers propagate steadily and localize in either time or space, in particular Akhmediev breather (AB)\cite{05,06} and Kuznetsov-Ma breather (KM)\cite{07}. AB breathes in space periodically and localize in time, while KM breathes in time and localize in space. Interestingly, by taking the breathing period of the two kinds of breather to infinity, rogue waves which are localized in both time and space may be obtained. In 1983\cite{08}, Peregrine first found a simple rational solution-Peregrine soliton, which was the limiting case of KM breather and especially considered as the rogue wave prototype\cite{09}.

 There have been many papers on rogue wave and semi-rational localized waves of single-component system, such as the nonlinear Schr\"{o}dinger(NLS) equation\cite{10,11,12}, the derivative NLS equation\cite{13,14,15}, the Davey-Stewartson equation\cite{16}, the Hirota equation\cite{17}, the Kundu-Eckhaus equation\cite{18}, the complex short pulse equation\cite{19}, the Sasa-Satsuma equation\cite{20}, and so on. While, a variety of complex system, such as Bose-Einstein condensates and nonlinaer optical fiber, usually involve more than one component\cite{21,22,23}. So, the discussion of localized waves in multicomponent coupled systems are meaningful and necessary. Some different solutions of  the mixed coupled NLS equation has been classficated in \cite{24}. The Maxwell-Bloch equtions \cite{25}in two-level optical medium have been solved to get breather, dark breather, rogue wave and dark rogue wave, meanwhile, performing the generalized Darboux transformation (DT) to these equations, one may acquire higher-order rogue waves and W-shaped solitons\cite{26}.

In this letter, enlightened by Baronio\cite{27} and Guo\cite{11,28}, we extend two-component coupled Hirota eauation\cite{29,30,31} to three-component one\cite{32}, and discuss localized nonlinear waves in this three-component coupled Hirota equation
\begin{equation}
\begin{array}{l}
iq_{1t}+\frac{1}{2}q_{1xx}+(|q_1|^2+|q_2|^2+|q_3|^2)q_1+i\epsilon[q_{1xxx}+3(2|q_1|^2+|q_2|^2+|q_3|^2)q_{1x}+3q_1(q_2^{*}q_{2x}+q_3^{*}q_{3x})]=0,\\
iq_{2t}+\frac{1}{2}q_{2xx}+(|q_1|^2+|q_2|^2+|q_3|^2)q_2+i\epsilon[q_{2xxx}+3(2|q_2|^2+|q_1|^2+|q_3|^2)q_{2x}+3q_2(q_1^{*}q_{1x}+q_3^{*}q_{3x})]=0,\\
iq_{3t}+\frac{1}{2}q_{3xx}+(|q_1|^2+|q_2|^2+|q_3|^2)q_3+i\epsilon[q_{3xxx}+3(2|q_3|^2+|q_1|^2+|q_3|^2)q_{3x}+3q_3(q_1^{*}q_{1x}+q_2^{*}q_{2x})]=0.
\end{array}
\end{equation}
Where $q_1(x,t)$, $q_2(x,t)$ and $q_3(x,t)$ are the complex envelops of three fields, each non-numeric subscripted variable stands for partial differentiation. Besides, $q_i^{*}(i=1,2,3)$ denotes the complex conjugate of $q_i$.  $\epsilon$  stands for the integrable perturbation of the coupled NLS equation\cite{28}, which is a small dimensionless real parameter. In the regime of ultra-short pulses, where the pulse lengths become comparable to the wavelength\cite{19,33}, the NLS equation becomes less accurate. To meet this requirement, one of the approaches is to add some higher-order dispersive terms\cite{34}. In this way, Tasgal and Potasek present the coupled Hirota equation\cite{36}. Besides, the coupled Hirota system includes third-order dispersion, self-steepening and delayed nonlinear response, thus, can be considered to be a more accurate prototype of the wave envolution in the real world.

In \cite{27} and \cite{28}, their methods can not obtain higher-order localized waves, so we construct a specifical vector solution of Lax pair of this equation in\cite{36}. Combing the generalized Darboux transformation and this specifical vector solution, the higher-order localized waves have been constructed in two-component coupled NLS equation. Several rational and semi-rational solutions are also obtained in coupled Hirota equation\cite{30,31}.The higher-order localized waves\cite{37} and four-petaled flower rogue wave\cite{38} are exhibited in three-component coupled NLS equation. Compared with the two-component coupled NLS equation, the three-component one has some new and appealing combination interactional solutions among these three component. So they are not just the same. One has constructed several new higher-order rogue wave structure in three-wave resonant interaction equation\cite{39}. Using similarity transformation technique, some localized matter waves in multicomponent Bose-Einstein condensates have been presented\cite{40}. Utilizing Painlev$\acute{e}$ analysis, the multi dark solition  of \emph{N}-component coupled Hirota equation\cite{32} has been discussed. In \cite{31}, the first- and second-order localized waves have been presented and their dynamic properties also have been discussed in detail of the two-component coupled Hirota equation. So, we will extend two-component coupled Hirota equation to three-component one, then, discuss diverse interactional solutions of this coupled system. In the paper, using our method, some meaningful results can be obtained in this three-component coupled Hirota equation. Besides, choosing the appropriate values of some free parameters in this semi-rational solution, several interesting dynamics of the interactional solutions are exhibited.

The paper is organized as follows. In section 2, the generalized Darboux transformation of Eq.(1) is constructed. In section 3, the first- and second-order localized waves are obtained respectively, and some interesting and appealing figures are also given. The last section contains several conclusions and discussions.

\section{Generalized Darboux transformation}

In this section, we reconstruct the following Lax pair with $4\times4$ matrixes of Eq.(1)\cite{31}, then give the generalized DT of this system,
\begin{eqnarray}
&&\Phi_{x}=U\Phi=(\lambda U_0+U_1)\Phi,\\
&&\Phi_{t}=V\Phi=(\lambda^3 V_0+\lambda^2 V_1+\lambda V_2+V_3)\Phi,
\end{eqnarray}
Where
\begin{gather*}
U_0=\frac{1}{12\epsilon}\begin{bmatrix}-2i&0&0&0\\0&i&0&0\\0&0&i&0\\0&0&0&i \end{bmatrix},\quad
U_1=\begin{bmatrix}0&-q_1&-q_2&-q_3\\ q_1^{*}&0&0&0\\ q_2^{*}&0&0&0\\ q_3^{*}&0&0&0 \end{bmatrix},\quad V_0=\frac{1}{16\epsilon}U_0,\quad V_1=\frac{1}{8\epsilon}U_0+\frac{1}{16\epsilon}U_1,
\end{gather*}

\begin{eqnarray*}
V_2=\frac{1}{4}\begin{bmatrix} ie&-\frac{q_1}{2\epsilon}-iq_{1x}&-\frac{q_2}{2\epsilon}-iq_{2x}&-\frac{q_3}{2\epsilon}-iq_{3x}\\
                               \frac{q_1^{*}}{2\epsilon}-iq_{1x}^{*}&-i|q_1|^2&-iq_1^{*}q_2&-iq_1^{*}q_3\\
                               \frac{q_2^{*}}{2\epsilon}-iq_{2x}^{*}&-iq_2^{*}q_1&-i|q_2|^2&-iq_2^{*}q_3\\
                               \frac{q_3^{*}}{2\epsilon}-iq_{3x}^{*}&-iq_3^{*}q_1&-iq_3^{*}q_2&-i|q_3|^2 \end{bmatrix},
\end{eqnarray*}
\begin{eqnarray*}
V_3=\begin{bmatrix} \epsilon(e_1+e_2+e_3)+\frac{i}{2}e & \epsilon e_4-\frac{i}{2}q_{1x} & \epsilon e_5-\frac{i}{2}q_{2x} & \epsilon e_6-\frac{i}{2}q_{3x}\\
                     -\epsilon e_4^{*}-\frac{i}{2}q_{1x}^{*} & -\epsilon e_1-\frac{i}{2}|q_1|^2 & \epsilon e_7-\frac{i}{2}q_1^{*}q_2 & \epsilon e_8-\frac{i}{2}q_1^{*}q_3\\
                     -\epsilon e_5^{*}-\frac{i}{2}q_{2x}^{*} & -\epsilon e_7^{*}-\frac{i}{2}q_2^{*}q_1 & -\epsilon e_2-\frac{i}{2}|q_2|^2 & \epsilon e_9-\frac{i}{2}q_2^{*}q_3\\
                     -\epsilon e_6^{*}-\frac{i}{2}q_{3x}^{*} & -\epsilon e_8^{*}-\frac{i}{2}q_3^{*}q_1 & -\epsilon e_9^{*}-\frac{i}{2}q_3^{*}q_2 & -\epsilon e_3-\frac{i}{2}|q_3|^2 \end{bmatrix},
 \end{eqnarray*}
with
\begin{eqnarray}
\nonumber && e=|q_1|^2+|q_2|^2+|q_3|^2,\quad \quad e_1=q_1q_{1x}^{*}-q_{1x}q_1^{*},\quad \quad e_2=q_2q_{2x}^{*}-q_{2x}q_2^{*},\\
\nonumber &&e_3=q_3q_{3x}^{*}-q_{3x}q_3^{*},\quad \quad e_4=q_{1xx}+2eq_1,\quad \quad e_5=q_{2xx}+2eq_2,\\
\nonumber &&e_6=q_{3xx}+2eq_3,\quad e_7=q_1^{*}q_{2x}-q_{1x}^{*}q_2,\quad e_8=q_1^{*}q_{3x}-q_{1x}^{*}q_3,\quad e_9=q_2^{*}q_{3x}-q_{2x}^{*}q_3.
\end{eqnarray}
Here, the column vector $\Phi=(\phi,\varphi,\chi,\psi)^{T}$ is eigenfunction and $\lambda$ is spectral parameter. Actually, Eq.(1) can be directly  figured out by the following compatibility condition $U_t-V_x+[U,V]=0$.

The Lax pair of Eq.(1) is standard AKNS (Ablowitz-Kaup-Newell-Segur) spectral problem, so, based on the DT of AKNS \cite{41} hierarchy, the generalized DT of Eq.(1) could be also constructed. However, in Eq.(2) and (3), $U$ and $V$ are all $4\times 4$ matrices, it is more complicated than $2\times2$ and $3\times3$ matrix to get a specifical vector solution of Lax pair.

Let $\Phi_1=(\phi_1,\varphi_1,\chi_1,\psi_1)^{T}$ be a special solution of Lax pair (2) and (3), with choosing the seed solution of Eq.(1) $q_1=q_1[0], q_2=q_2[0], q_3=q_3[0]$, at $\lambda=\lambda_1$. Then, we can get the classical DT of Eq.(1)
\begin{eqnarray}
&&\Phi_1=T[1]\Phi,\quad T[1]=\lambda I-H[0]\Lambda_1H[0]=(\lambda-\lambda_1^{*})I+(\lambda_1^{*}-\lambda_1)\dfrac{\Phi_1[0]\Phi_1[0]^{\dag}}{\Phi_1[0]^{\dag}\Phi_1[0]},\\ &&q_1[1]=q_1[0]+i(\lambda_1-\lambda_1^{*})\dfrac{\phi_1[0]\varphi_1[0]^{*}}{4\epsilon(|\phi_1[0]|^2+|\varphi_1[0]|^2+|\chi_1[0]|^2+|\psi_1[0]|^2)},\\
&&q_2[1]=q_2[0]+i(\lambda_1-\lambda_1^{*})\dfrac{\phi_1[0]\chi_1[0]^{*}}{4\epsilon(|\phi_1[0]|^2+|\varphi_1[0]|^2+|\chi_1[0]|^2+|\psi_1[0]|^2)},\\
&&q_3[1]=q_3[0]+i(\lambda_1-\lambda_1^{*})\dfrac{\phi_1[0]\psi_1[0]^{*}}{4\epsilon(|\phi_1[0]|^2+|\varphi_1[0]|^2+|\chi_1[0]|^2+|\psi_1[0]|^2)},
\end{eqnarray}
Where $(\phi_1[0],\varphi_1[0],\chi_1[0],\psi_1[0])^{T}=(\phi_1,\varphi_1,\chi_1,\psi_1)^{T}=\Phi_1[0]$, $\dag$ denotes transposed and conjugate operation of a matrix ( or a vector), $I$ is the $4\times 4$ identity matrix,
\begin{eqnarray}
H[0]=\begin{bmatrix} \phi_1[0]&\varphi_1[0]^{*}&\psi_1[0]^{*}&0\\ \varphi_1[0]&-\phi_1[0]^{*}&0&0\\ \chi_1[0]&0&0&\psi_1[0]^{*}\\ \psi_1[0]&0&-\phi_1[0]^{*}&-\chi_1[0]^{*} \end{bmatrix},\quad
\Lambda_1=\begin{bmatrix} \lambda_1&0&0&0\\ 0&\lambda_1^{*}&0&0\\ 0&0&\lambda_1^{*}&0\\ 0&0&0&\lambda_1^{*} \end{bmatrix}.
\end{eqnarray}

In the following, according to the above classical DT of Eq.(1), let $\Phi_1=(\phi_1,\varphi_1,\chi_1,\psi_1)^{T}=\Phi_1(\lambda_1+\delta)$ is a special solution of Eq.(2) and (3) with $q_1=q_1[0], q_2=q_2[0], q_3=q_3[0]$ and $\lambda=\lambda_1+\delta$, then $\Phi_1$ can be expanded as the Taylor series at $\delta=0$
\begin{eqnarray}
\Phi_1=\Phi_1^{[0]}+\Phi_1^{[1]}\delta+\Phi_1^{[2]}\delta^2+\cdots+\Phi_1^{[N]}\delta^N+\cdots,
\end{eqnarray}
Where
\begin{center}
$\Phi_1^{[l]}=(\phi_1^{[l]},\varphi_1^{[l]},\chi_1^{[l]},\psi_1^{[l]})^T$,\quad $\Phi_1^{[l]}=\dfrac{1}{l!}\dfrac{\partial^l \Phi_1}{\partial \delta^l}|_{\delta=0}\quad(l=0,1,2,3\cdots)$.
\end{center}

It can be easily figured out that $\Phi_1[0]=\Phi_1^{[0]}$ is a particular solution of Eqs.(2) and (3) with the above seed solution of Eq.(1) and at $\lambda=\lambda_1$ from the above process. So, we can directly give the first-step generalized DT through Eq.(4)-(7).

1. The first-step generalized DT
\begin{eqnarray}
&&\Phi_1=T[1]\Phi,\quad T[1]=\lambda I-H[0]\Lambda_1H[0]^{-1},\\
&&q_1[1]=q_1[0]+i(\lambda_1-\lambda_1^{*})\dfrac{\phi_1[0]\varphi_1[0]^{*}}{4\epsilon(|\phi_1[0]|^2+|\varphi_1[0]|^2+|\chi_1[0]|^2+|\psi_1[0]|^2)},\\
&&q_2[1]=q_2[0]+i(\lambda_1-\lambda_1^{*})\dfrac{\phi_1[0]\chi_1[0]^{*}}{4\epsilon(|\phi_1[0]|^2+|\varphi_1[0]|^2+|\chi_1[0]|^2+|\psi_1[0]|^2)},\\
&&q_3[1]=q_3[0]+i(\lambda_1-\lambda_1^{*})\dfrac{\phi_1[0]\psi_1[0]^{*}}{4\epsilon(|\phi_1[0]|^2+|\varphi_1[0]|^2+|\chi_1[0]|^2+|\psi_1[0]|^2)},
\end{eqnarray}
Where $\Phi_1[0]=\Phi_1^{[0]}=(\phi_1[0],\varphi_1[0],\chi_1[0],\psi_1[0])^{T}$, meanwhile, the explicit expressions of $H[0]$ and $\Lambda_1$ are given by Eq.(8).

2. The second-step generalized DT

Choosing the seed solution of Eq.(1) as $q_1=q_1[1], q_2=q_2[1], q_3=q_3[1]$ and $\lambda=\lambda_1+\delta$, then $T[1]\Phi_1$ is the solution of Eq.(2) and (3). So, we consider the following limit
\begin{eqnarray*}
\lim\limits_{\delta\rightarrow{0}}\dfrac{T[1]|_{\lambda=\lambda_1+\delta}\Phi_1}{\delta}=\lim\limits_{\delta\rightarrow{0}}\dfrac{(\delta+T_1[1])\Phi_1}{\delta}
=\Phi_1^{[0]}+T_1[1]\Phi_1^{[1]}\equiv\Phi_1[1],
\end{eqnarray*}
then
\begin{eqnarray}
&&\Phi_2=T[2]T[1]\Phi,\quad T[2]=\lambda I-H[1]\Lambda_2H[1]^{-1},\\
&&q_1[2]=q_1[1]+i(\lambda_1-\lambda_1^{*})\dfrac{\phi_1[1]\varphi_1[1]^{*}}{4\epsilon(|\phi_1[1]|^2+|\varphi_1[1]|^2+|\chi_1[1]|^2+|\psi_1[1]|^2)},\\
&&q_2[2]=q_2[1]+i(\lambda_1-\lambda_1^{*})\dfrac{\phi_1[1]\chi_1[1]^{*}}{4\epsilon(|\phi_1[1]|^2+|\varphi_1[1]|^2+|\chi_1[1]|^2+|\psi_1[1]|^2)},\\
&&q_3[2]=q_3[1]+i(\lambda_1-\lambda_1^{*})\dfrac{\phi_1[1]\psi_1[1]^{*}}{4\epsilon(|\phi_1[1]|^2+|\varphi_1[1]|^2+|\chi_1[1]|^2+|\psi_1[1]|^2)},
\end{eqnarray}
Where $\Phi_1[1]=(\phi_1[1],\varphi_1[1],\chi_1[1],\psi_1[1])^{T}$, $\Lambda_2=\Lambda_1$ and $T_1[1]=\lambda_1 I-H[0]\Lambda_1H[0]^{-1}$,
\begin{eqnarray*}
H[1]=\begin{bmatrix} \phi_1[1]&\varphi_1[1]^{*}&\psi_1[1]^{*}&0\\ \varphi_1[1]&-\phi_1[1]^{*}&0&0\\ \chi_1[1]&0&0&\psi_1[1]^{*}\\ \psi_1[1]&0&-\phi_1[1]^{*}&-\chi_1[1]^{*} \end{bmatrix}.
\end{eqnarray*}

3. The third-step generalized DT

In the same way, the following kind of limit will be constructed
\begin{eqnarray}
\nonumber\lim\limits_{\delta\rightarrow{0}}\dfrac{(T[2]T[1])|_{\lambda=\lambda_1+\delta}\Phi_1}{\delta^2}=\lim\limits_{\delta\rightarrow{0}}\dfrac{(\delta+T_1[2])(\delta+T_1[1])\Phi_1}{\delta^2}\\
\nonumber\quad \quad=\Phi_1^{[0]}+(T_1[1]+T_1[2])\Phi_1^{[1]}+T_1[2]T_1[1]\Phi_1^{[2]}\equiv\Phi_1[2],
\end{eqnarray}
then, the specifical solution of Lax pair (2) and (3) can be solved at $q_1=q_1[2], q_2=q_2[2], q_3[2]$ with the spectral parameter $\lambda=\lambda_1$. Certainly, these two identities could be presented
\begin{eqnarray}
\nonumber&&T_1[1]\Phi_1^{[0]}=0,\quad \quad T_1[2](\Phi_1^{[0]}+T_1[1]\Phi_1^{[1]})=0.\\
&&\Phi_3=T[3]T[2]T[1]\Phi,\quad T[3]=\lambda I-H[2]\Lambda_3H[2]^{-1},\\
&&q_1[3]=q_1[2]+i(\lambda_1-\lambda_1^{*})\dfrac{\phi_1[2]\varphi_1[2]^{*}}{4\epsilon(|\phi_1[2]|^2+|\varphi_1[2]|^2+|\chi_1[2]|^2+|\psi_1[2]|^2)},\\
&&q_2[3]=q_2[2]+i(\lambda_1-\lambda_1^{*})\dfrac{\phi_1[2]\chi_1[2]^{*}}{4\epsilon(|\phi_1[2]|^2+|\varphi_1[2]|^2+|\chi_1[2]|^2+|\psi_1[2]|^2)},\\
&&q_3[3]=q_3[2]+i(\lambda_1-\lambda_1^{*})\dfrac{\phi_1[2]\psi_1[2]^{*}}{4\epsilon(|\phi_1[2]|^2+|\varphi_1[2]|^2+|\chi_1[2]|^2+|\psi_1[2]|^2)},
\end{eqnarray}
Where $\Phi_1[2]=(\phi_1[2],\varphi_1[2],\chi_1[2],\psi_1[2])^{T}$, $\Lambda_3=\Lambda_1$ and $T_1[2]=\lambda_1 I-H[1]\Lambda_2H[1]^{-1}$,
\begin{eqnarray*}
H[2]=\begin{bmatrix} \phi_1[2]&\varphi_1[2]^{*}&\psi_1[2]^{*}&0\\ \varphi_1[2]&-\phi_1[2]^{*}&0&0\\ \chi_1[2]&0&0&\psi_1[2]^{*}\\ \psi_1[2]&0&-\phi_1[2]^{*}&-\chi_1[2]^{*} \end{bmatrix}.
\end{eqnarray*}

4. The \emph{N}-step generalized DT

Iterating the above procedures, thus, the \emph{N}-generalized DT of Eq.(1) can be defined as the following form:
\begin{eqnarray}
&&\Phi[N]=T[N]T[N-1]\cdots T[1]\Phi,\quad \quad T[N]=\lambda I-H[N-1]\Lambda_NH[N-1]^{-1},\\
\nonumber&&\Phi_1[N-1]=\Phi_1^{[0]}+\sum^{N-1}_{l=1}T_1[l]\Phi_1^{[1]}+\sum^{N-1}_{l=1}\sum^{l-1}_{k=1}T_1[l]T_1[k]\Phi_1^{[2]}+\cdots+T_1[N-1]T_1[N-2]\\
&&\hspace{2.5cm}\cdots T_1[1]\Phi_1^{[N-1]},\\
&&q_1[N]=q_1[N-1]+\dfrac{i(\lambda_1-\lambda_1^{*})\phi_N[N-1]\varphi_N[N-1]^{*}}{4\epsilon(|\phi_N[N-1]|^2+|\varphi_N[N-1]|^2+|\chi_N[N-1]|^2+|\psi_N[N-1]|^2)},\\
&&q_2[N]=q_2[N-1]+\dfrac{i(\lambda_1-\lambda_1^{*})\phi_N[N-1]\chi_N[N-1]^{*}}{4\epsilon(|\phi_N[N-1]|^2+|\varphi_N[N-1]|^2+|\chi_N[N-1]|^2+|\psi_N[N-1]|^2)},\\
&&q_3[N]=q_3[N-1]+\dfrac{i(\lambda_1-\lambda_1^{*})\phi_N[N-1]\psi_N[N-1]^{*}}{4\epsilon(|\phi_N[N-1]|^2+|\varphi_N[N-1]|^2+|\chi_N[N-1]|^2+|\psi_N[N-1]|^2)},
\end{eqnarray}
Where
\begin{eqnarray*}
&&H[l-1]=\begin{bmatrix} \phi_1[l-1]&\varphi_1[l-1]^{*}&\psi_1[l-1]^{*}&0\\ \varphi_1[l-1]&-\phi_1[l-1]^{*}&0&0\\ \chi_1[l-1]&0&0&\psi_1[l-1]^{*}\\ \psi_1[l-1]&0&-\phi_1[l-1]^{*}&-\chi_1[l-1]^{*} \end{bmatrix}, \Lambda_l=\begin{bmatrix} \lambda_1&0&0&0\\ 0&\lambda_1^{*}&0&0\\ 0&0&\lambda_1^{*}&0\\ 0&0&0&\lambda_1^{*} \end{bmatrix}(1\leq l \leq N),\\
&&\Phi_1[N-1]=(\phi_1[N-1],\varphi_1[N-1],\chi_1[N-1],\psi_1[N-1])^{T}~(N\geq1),\quad T_1[l]=\lambda_1 I-H[l-1]\Lambda_lH[l-1]^{-1}.
\end{eqnarray*}
Here, we can find that Eq.(24)-(26) give rise to the \emph{N}th-order localized waves in Eq.(1) theoretically. In order to avoid calculating the determinant of higher order matrix in a cumbersome way. Instead of Crum theorem\cite{42}, the iterative algorithm should be chosen in this paper.

\section{Localized nonlinear wave solutions}
In this section, we present  the localized nonlinear wave solutions of this three-component coupled Hirota system\cite{32}, and demonstrate the dynamics analysis of these nonlinear waves in detail.
\subsection{The first-order localized nonlinear waves}
We begin with the plane wave solution of Eq.(1)\cite{28,32}
\begin{eqnarray}
q_1[0]=d_1e^{i\theta},\quad \quad q_2[0]=d_2e^{i\theta}, \quad \quad q_3[0]=d_3e^{i\theta},
\end{eqnarray}
Where, $\theta=(d_1^2+d_2^2+d_3^2)t$, and $d_1, d_2, d_3$ are three arbitrary real constants, which are the backgrounds emerging these localized nonlinear waves. Conveniently, the above seed solutions are chosen periodically in time variable $t$ without depending on space variable $x$. Then the peculiar vector solution of Lax pair (2) and (3) with $\lambda$ at $q_1=q_1[0], q_2=q_2[0]$ and $q_3=q_3[0]$, can be constructed
\begin{eqnarray}
\Phi_1=\begin{pmatrix}(c_1e^{M_1+M_2}-c_2e^{M_1-M_2})e^{\tfrac{i\theta}{2}}\\\rho_1(c_2e^{M_1+M_2}-c_1e^{M_1-M_2})e^{-\tfrac{i\theta}{2}}-(\alpha d_2+\beta d_3)e^{M_3}\\ \rho_2(c_2e^{M_1+M_2}-c_1e^{M_1-M_2})e^{-\tfrac{i\theta}{2}}+\alpha d_1e^{M_3}\\ \rho_3(c_2e^{M_1+M_2}-c_1e^{M_1-M_2})e^{-\tfrac{i\theta}{2}}+\beta d_1e^{M_3} \end{pmatrix},
\end{eqnarray}
where
\begin{eqnarray}
\nonumber &&c_1=\dfrac{\left(\lambda-\sqrt{\lambda^2+64\epsilon^2(d_1^2+d_2^2+d_3^2)}\right)^{\tfrac{1}{2}}}{\sqrt{\lambda^2+64\epsilon^2(d_1^2+d_2^2+d_3^2)}}, \quad c_2=\dfrac{\left(\lambda+\sqrt{\lambda^2+64\epsilon^2(d_1^2+d_2^2+d_3^2)}\right)^{\tfrac{1}{2}}}{\sqrt{\lambda^2+64\epsilon^2(d_1^2+d_2^2+d_3^2)}},\\
\nonumber && \rho_1=\dfrac{d_1}{\sqrt{d_1^2+d_2^2+d_3^2}},\quad \quad  \rho_2=\dfrac{d_2}{\sqrt{d_1^2+d_2^2+d_3^2}},\quad \quad \rho_3=\dfrac{d_3}{\sqrt{d_1^2+d_2^2+d_3^2}},\\
\nonumber && M_1=-\dfrac{i\lambda}{384\epsilon^2}[16\epsilon x+\lambda(\lambda+2)t],\quad \quad M_3=\dfrac{i\lambda}{192\epsilon^2}[16\epsilon x+\lambda(\lambda+2)t],\\
\nonumber && M_2=\dfrac{i}{128\epsilon^2}\sqrt{\lambda^2+64\epsilon^2(d_1^2+d_2^2+d_3^2)}[16\epsilon x +\lambda(\lambda+2)t-32\epsilon^2(d_1^2+d_2^2+d_3^2)+\sum_{k=1}^{N}s_kf^{2k}].
\end{eqnarray}
Here, $s_k=m_k+in_k(1\leq k\leq N)$, and $m_k, n_k, \alpha, \beta$ all could be chosen as real constants freely. Setting $\tau=d_1^2+d_2^2+d_3^2$ and choosing the spectral parameter $\lambda=8i\sqrt{\tau}\epsilon(1+f^2)$ with a small real parameter $f$, we can get the Taylor expansion of the vector function $\Phi_1$ at $f=0$
\begin{eqnarray}
\Phi_1(f)=\Phi_1^{[0]}+\Phi_1^{[1]}f^2+\Phi_1^{[2]}f^4+\Phi_1^{[3]}f^6+\cdots,
\end{eqnarray}
where, $\Phi_1^{[k]}=(\phi_1^{[k]},\varphi_1^{[k]},\chi_1^{[k]},\psi_1^{[k]})^{T}=\dfrac{\partial^{2k} \Phi_1}{\partial f^{2k}}|_{f=0}~(1\leq k\leq N)$,

\begin{eqnarray}
\nonumber && \phi_1^{[0]}=\dfrac{(i-1)[2\sqrt{\tau}(x-6\tau\epsilon t)+2i\tau t+1]}{4\sqrt{\epsilon}\tau^{\tfrac{1}{4}}}e^{\xi_1},\\
 \nonumber && \varphi_1^{[0]}=\dfrac{(i-1)d_1[2\sqrt{\tau}(x-6\tau\epsilon t)+2i\tau t-1]}{4\sqrt{\epsilon}\tau^{\tfrac{3}{4}}}e^{\xi_2}-(\alpha d_2+\beta d_3)e^{\xi_3},\\
\nonumber  && \chi_1^{[0]}=\dfrac{(i-1)d_2[2\sqrt{\tau}(x-6\tau\epsilon t)+2i\tau t-1]}{4\sqrt{\epsilon}\tau^{\tfrac{3}{4}}}e^{\xi_2}+\alpha d_1 e^{\xi_3},\\
 \nonumber && \psi_1^{[0]}=\dfrac{(i-1)d_3[2\sqrt{\tau}(x-6\tau\epsilon t)+2i\tau t-1]}{4\sqrt{\epsilon}\tau^{\tfrac{3}{4}}}e^{\xi_2}+\beta d_1e^{\xi_3},\\
 \nonumber && \phi_1^{[1]}=\dfrac{1-i}{96\tau^{\tfrac{1}{4}}\epsilon^{\tfrac{3}{2}}}[-16\tau^{\tfrac{3}{2}}\epsilon x^3+288\tau^{\tfrac{5}{2}}\epsilon^2x^2t+48\tau^{\tfrac{5}{2}}\epsilon x t^2-1728\tau^{\tfrac{7}{2}}\epsilon^3 x t^2+3456\tau^{\tfrac{9}{2}}\epsilon^4t^3\\
\nonumber && \hspace{0.9cm} -288\tau^{\tfrac{7}{2}}\epsilon^2t^3-40\epsilon\tau x^2+576\epsilon^2\tau^2xt-2016\epsilon^3\tau^3t^2+56\tau^2\epsilon t^2-20\sqrt{\tau}\epsilon x+552\tau^{\tfrac{3}{2}}\epsilon^2t\\
\nonumber && \hspace{0.9cm}-3\sqrt{\tau}m_1+6\epsilon +i(-1728\tau^4\epsilon^3t^3+576\tau^3\epsilon^2xt^2+16\tau^3\epsilon t^3
-48\tau^2\epsilon x^2t\\
\nonumber && \hspace{0.9cm}-96\tau^{\tfrac{3}{2}}\epsilon xt+672\tau^{\tfrac{5}{2}}\epsilon^2t^2-76\tau\epsilon t-3\sqrt{\tau}n_1)]e^{\xi_1},\\
 \nonumber && \varphi_1^{[1]}=-\frac{2}{3}(12\tau^{\frac{3}{2}}\epsilon t-2i\tau t-\sqrt{\tau}x)(\alpha d_2+\beta d_3)e^{\xi_3}+\dfrac{d_1\Omega}{96\epsilon^{\tfrac{3}{2}}\tau^{\tfrac{3}{4}}}e^{\xi_2},\\
\nonumber && \chi_1^{[1]}=\frac{2}{3}\alpha d_1(12\tau^{\tfrac{3}{2}}\epsilon t-2i\tau t-\sqrt{\tau}x)e^{\xi_3}-\dfrac{d_2\Omega}{96\epsilon^{\tfrac{3}{2}}\tau^{\tfrac{3}{4}}}e^{\xi_2},\\
 \nonumber && \psi_1^{[1]}=\frac{2}{3}\beta d_1(12\tau^{\tfrac{3}{2}}\epsilon t-2i\tau t-\sqrt{\tau}x)e^{\xi_3}-\dfrac{d_3\Omega}{96\epsilon^{\tfrac{3}{2}}\tau^{\tfrac{3}{4}}}e^{\xi_2},\\
 \nonumber &&  \cdots \cdots
\end{eqnarray}
with
\begin{eqnarray}
\nonumber && \xi_1=\frac{1}{3}\sqrt{\tau}x+\tau(\frac{5}{6}i-\frac{4}{3}\epsilon \sqrt{\tau})t,\\
\nonumber && \xi_2=\frac{1}{3}\sqrt{\tau}x-\tau(\frac{1}{6}i+\frac{4}{3}\epsilon \tau)t,\\
\nonumber &&\xi_3= -\frac{2}{3}\sqrt{\tau}x-\tau(\frac{2}{3}i-\frac{8}{3}\sqrt{\tau}\epsilon)t,\\
\nonumber && \Omega=(i-1)[-16\tau^{\tfrac{3}{2}}\epsilon x^3+288\tau^{\tfrac{5}{2}}\epsilon^2x^2t+48\tau^{\tfrac{5}{2}}\epsilon x t^2-1728\tau^{\tfrac{7}{2}}\epsilon^3xt^2+3456\tau^{\tfrac{9}{2}}\epsilon^4t^3
-288\tau^{\tfrac{7}{2}}\epsilon^2t^3\\
\nonumber &&\hspace{0.9cm}+8\epsilon\tau x^2+8\tau^2\epsilon t^2-288\epsilon^3\tau^3t^2-4\sqrt{\tau}\epsilon x+360\tau^{\tfrac{3}{2}}\epsilon^2 t-3 \sqrt{\tau}m_1-6\epsilon i(-1728\tau^4\epsilon^3t^3\\
\nonumber&&\hspace{0.9cm}+576\tau^3\epsilon^2xt^2+16\tau^3\epsilon t^3-48\tau^2\epsilon x^2t-44\tau\epsilon t+96\epsilon^2\tau^{\tfrac{5}{2}}t^2-3\sqrt{\tau}n_1).
\end{eqnarray}

It can be straightforward to calculate that the vector function $\Phi_1^{[0]}$ is a solution of Lax pair (2) and (3), at $\lambda=\lambda_1=8i\sqrt{\tau}\epsilon$ and $q_1=q_1[0], q_2=q_2[0], q_3=q_3[0]$. Hence, through Eqs.(11)-(13), we can get
\begin{eqnarray}
&& q_1[1]=d_1e^{i\theta}+\dfrac{2[d_1\tau^{\tfrac{1}{4}}\sqrt{\epsilon}F_1e^{i\theta}-2\tau \epsilon(\alpha d_2+\beta d_3) F_2e^{\eta_1}]}{\tau^{\tfrac{1}{4}}\sqrt{\epsilon}(4\sqrt{\tau}\epsilon G_1e^{\eta_2}+G_2)},\\
&& q_2[1]=d_2e^{i\theta}+\dfrac{2(d_2\tau^{\tfrac{1}{4}}\sqrt{\epsilon}F_1 e^{i\theta}+2\alpha d_1\tau \epsilon F_2e^{\eta_1})}{\tau^{\tfrac{1}{4}}\sqrt{\epsilon}(4\sqrt{\tau}\epsilon G_1e^{\eta_2}+G_2)},\\
&& q_3[1]=d_3e^{i\theta}+\dfrac{2(d_3\tau^{\tfrac{1}{4}}\sqrt{\epsilon}F_1 e^{i\theta}+2\beta d_1\tau \epsilon F_2e^{\eta_1})}{\tau^{\tfrac{1}{4}}\sqrt{\epsilon}(4\sqrt{\tau}\epsilon G_1e^{\eta_2}+G_2)},
\end{eqnarray}
where
\begin{eqnarray*}
&& F_1=-4\tau x^2+48\tau^2\epsilon x t-4\tau^2t^2-144\tau^3\epsilon^2t^2+4i\tau t+1,\\
&&F_2=(-1+i)(12\tau^{\tfrac{3}{2}}\epsilon t-2i\tau t-2\sqrt{\tau}x-1),\\
&&G_1=\alpha^2d_1^2+\alpha^2d_2^2+2\alpha\beta d_2d_3+\beta^2d_1^2+\beta^2d_3^2,\\
&&G_2=4\tau x^2-48\tau^2\epsilon x t+144\tau^3\epsilon^2t^2+4\tau^2t^2+1.
\end{eqnarray*}

The correctness of the above Eqs.(30)-(32) have been directly verified by putting them back into Eq.(1). At this moment, we obtain the first-order localized nonlinear waves of Eq.(1) with five free parameters $d_1, d_2, d_3, \alpha, \beta$. Where, $d_1, d_2$ and $d_3$ determine the background where the different localized waves emerge, and $\alpha , \beta $ play the important role in controlling dynamics of these nonlinear waves. Then, we will discuss dynamics properties of these solutions in six different cases.

(\textrm{i}) When $\alpha=\beta=0$, $q_1, q_2$ and $q_3$ are all proportional to each other, and we can obviously find they are the first-order rogue wave. Besides, these three component have the similar structure, and they are the same as the one-component and two-component cases from Fig.1
%

(\textrm{ii}) When one of the two parameters $\alpha, \beta$ is zero and $d_i\neq0~(i=1,2,3)$, without loss of generality, we choose $\alpha=0, \beta\neq0$. We can see the combination from Fig.2, $q_1$ and $q_3$ are all first-order RW and one-breather, $q_2$ is first-order RW and one-dark soliton. By increasing the absolute value $\beta$, the phenomenon that the nonlinear waves merge with each other distinctly can be shown in Fig.3.
%
%

(\textrm{iii}) When $\alpha=0,\beta\neq0$ and $d_1\neq0, d_2\neq0,d_3=0$, we can get the combination that $q_1$ and $q_2$ are all the first-order RW and one-dark soliton, $q_3$ is the first-order RW and one-bright soliton in Fig.4. Owing to the zero-amplitude plane which emerges the RW, we can not observe this RW clearly in Fig.4(c). In the same way, increasing the absolute values $\alpha$ and $\beta$, the character that these nonlinear waves integrate with each other can be shown in Fig.5.

%

(\textrm{iv}) When $\alpha \neq 0, \beta \neq0$, and $d_1\neq0, d_2\neq0, d_3=0$, the combination of  these three components is that $q_1$ and $q_2$ are all the first-order RW and one-breather, $q_3$ is the first-order RW and one-bright soliton in Fig.6. While considering the zero-amplitude background crest, the rogue wave in $q_3$ is difficult to observe owing to its small amplitude. And the more obvious interaction among different nonlinear waves can be found in Fig.7.
%

(\textrm{v}) When $\alpha \neq0, \beta \neq 0$, and $d_1\neq 0, d_2=d_3=0$, we can get a new combination among these three components $q_1$, $q_2$ and $q_3$. In Fig.8, the phenomenon, $q_1$ is the first-order RW and one-dark soliton, $q_2$ and $q_3$ are all the first-order RW and one-bright soliton, is exhibited. The maximum amplitude of first-order rogue wave in Fig.8(b) and (c) are very small and difficult to be find, because they appear at the zero-amplitude background crest. Similarly, we can see that these nonlinear waves emerge with each other distinctly from Fig.9. With respect to Fig.9, we give the explicit collision processes between one-dark soliton and a rogue wave, one-bright soliton and a rogue wave respectively in Fig.10. We can find that, in Fig.10, one-dark soliton and one-bright soliton are propagating along the positive direction of $x$-axis, when $t=0$, the first-order RW suddenly appears and these nonlinear waves interact with each other. The next moment, the first-order RW disappears without a trace and the solitons which maintain their identity continue to spread after this collision.

(\textrm{vi}) When $\alpha \neq0, \beta \neq0$, and $d_i~(i=1,2,3)$ are all not zero, the three component $q_1, q_2,$ and $q_3$ are all the first-order RW and one-breather in Fig.11. In the same way,increasing the absolute values of $\alpha$ and $\beta$, we can see the  RW integrates with one-breather distinctively in $q_1$, $q_2$ and $q_3$ from Fig.12.

Here, we give a simple classification about values of these five parameters $\alpha, \beta$ and $d_i~(i=1,2,3)$ corresponding to different types of the first-order nonlinear wave solutions of Eq.(1). These following nonlinear waves can be obtained by choosing appropriate values in these five parameters.

\textbf{Case} 1: When $\alpha=\beta=0$, these solutions $q_i~(i=1,2,3)$ are all the first-order rogue wave.

\textbf{Case} 2: One of the two parameters $\alpha$ and $\beta$ is zero, without lose of generality, we take the case of $\alpha=0, \beta \neq 0$ into account. This classification is shown in Table 1.
\begin{center}
{\footnotesize{Table 1.}Classification of the first-order localized waves solutions generated by the first-step generalized DT\\
\vspace{2mm}
\begin{tabular}{cccc}
  \hline
  $d_i$ & $q_1$ & $q_2$ & $q_3$ \\
  \hline
  $d_1\neq0, d_2=d_3=0$& {RW and one-dark soliton}  & {0} & {RW and one-bright soliton} \\
  $d_1=0,d_2\neq0, d_3=0$& {0} & {RW} & {0} \\
  $d_1=0,d_2=0,d_3\neq0$& {RW and one-bright soliton} & {0} & {RW and one-dark soliton} \\
  $d_1\neq0, d_2=0, d_3\neq0$&{RW and one-breather} &{0} &{RW and one-breather}\\
  $d_1\neq0, d_2\neq0,d_3=0$&{RW and one-dark soliton}&{RW and one-dark soliton}&{RW and one-bright soliton}\\
  $d_1=0,d_2\neq0,d_3\neq0$&{RW and one-bright soliton}&{RW and one-dark soliton}&{RW and one-dark soliton}\\
  $d_1\neq0,d_2\neq0,d_3\neq0$&{RW and one-breather}&{RW and one-dark soliton}&{RW and one-breather}\\
  \hline
\end{tabular}}
\end{center}
\vspace{2mm}

\textbf{Case} 3:When $\alpha\neq0, \beta\neq0$, the classification is shown in Table 2.

\begin{center}
{\footnotesize{Table 2.}Classification of the first-order localized waves solutions generated by the first-step generalized DT\\
\vspace{2mm}
\begin{tabular}{cccc}
  \hline
  $d_i$ & $q_1$ & $q_2$ & $q_3$ \\
  \hline
  $d_1\neq0, d_2=d_3=0$& {RW and one-dark soliton}  & {RW and one-bright soliton} & {RW and one-bright soliton} \\
  $d_1=0,d_2\neq0, d_3=0$& {RW and one-bright soliton} & {RW and one-dark soliton} & {0} \\
  $d_1=0,d_2=0,d_3\neq0$& {RW and one-bright soliton} & {0} & {RW and one-dark soliton} \\
  $d_1\neq0, d_2=0, d_3\neq0$&{RW and one-breather} &{RW and one-bright soliton} &{RW and one-breather}\\
  $d_1\neq0, d_2\neq0,d_3=0$&{RW and one-breather}&{RW and one-breather}&{RW and one-bright soliton}\\
  $d_1=0,d_2\neq0,d_3\neq0$&{RW and one-bright soliton}&{RW and one-dark soliton}&{RW and one-dark soliton}\\
  $d_1\neq0,d_2\neq0,d_3\neq0$&{RW and one-breather}&{RW and one-breather}&{RW and one-breather}\\
  \hline
\end{tabular}}
\end{center}

 Instead of considering various arrangements of the three potential functions $q_1,q_2$ and $q_3$, we consider the same combination as the same type solution. In other words, case 1 is that $q_1$ and $q_2$ are all RW and one-dark soliton, $q_3$ is RW and one-bright soliton; and case 2 is that $q_1$ is RW and one-bright soliton, $q_2$ and $q_3$ are all RW and one-dark soliton, in our opinion, these two cases is the same type solution. So, we can get six types first-order interactional solutions using our method in this paper.

\begin{center}
{\footnotesize{Table 3.} Six types first-order localized nonlinear waves\\
\vspace{2mm}
\begin{tabular}{cccc}
  \hline
  Types & $q_i~(i=1,2,3)$\\
  \hline
  Type 1& three potential functions are all first-order RW \\
  Type 2& the two potential functions are RW and one-breather, and another one is RW and one-dark soliton \\
  Type 3& the two potential functions are RW and one-dark soliton, and another one is RW and one-bright soliton\\
  Type 4& the two potential functions are RW and one-breather, and another one is RW and one-bright soliton\\
  Type 5& the two potential functions are RW and one-bright soliton, and another one is RW and one-dark soliton\\
  Type 6& three potential functions are all RW and one-breather\\
  \hline
\end{tabular}}
\end{center}
\subsection{The second-order localized nonlinear waves}
In this section, we consider the following limit:
\begin{eqnarray}
\nonumber &&\lim\limits_{f\rightarrow{0}}\dfrac{T[1]|_{\lambda=8i\sqrt{\tau}\epsilon(1+f^2)}\Phi_1}{f^2}=\lim\limits_{f\rightarrow{0}}\dfrac{(8i\sqrt{\tau}\epsilon f^2+T_1[1])\Phi_1}{f^2}\\
&&\hspace{4.5cm} =8i\sqrt{\tau} \epsilon \Phi_1^{[0]}+T_1[1]\Phi_1^{[1]}\equiv\Phi_1[1],
\end{eqnarray}
\begin{eqnarray}
\nonumber && T_1[1]=\lambda_1I-H[0]\Lambda_1H[0]^{-1}=(\lambda_1-\lambda_1^{*})I-(\lambda_1-\lambda_1^{*})\dfrac{\Phi_1[0]\Phi_1[0]^{\dag}}{\Phi_1[0]^{\dag}\Phi_1[0]}\\
\nonumber && \hspace{5.2cm}=(\lambda_1-\lambda_1^{*})I-(\lambda_1-\lambda_1^{*})\dfrac{\Phi_1^{[0]}\Phi_1^{[0]\dag}}{\Phi_1^{[0]\dag}\Phi_1^{[0]}},\\
 && \hspace{5.2cm}=16i\sqrt{\tau}\epsilon(I-\dfrac{\Phi_1^{[0]}\Phi_1^{[0]\dag}}{\Phi_1^{[0]\dag}\Phi_1^{[0]}}),
\end{eqnarray}
where $\Phi_1^{[0]}=\dfrac{\partial^{0} \Phi_1}{\partial f^0}|_{f=0}$, $\Phi_1^{[1]}=\dfrac{\partial^{2} \Phi_1}{\partial f^2}|_{f=0}$. We can arrive at a specific vector solution of Lax pair (2) and (3) with $q_1=q_1[1], q_2=q_2[1], q_3=q_3[1]$, at $\lambda=\lambda_1=8i\sqrt{\tau}\epsilon$. Through Eqs.(15)-(17), the concrete expressions of the second-order localized nonlinear waves can be figured out. However, this explicit expressions of $q_1[2], q_2[2]$ and $q_3[2]$ are very tedious and complicated, we only give their expressions of the simplest case $\alpha=\beta=0$ in the following content. For other cases, we omit writing down their expressions for the complexity, while, the dynamics properties of these solutions are discussed in detail. Besides, the validity of the expressions of $q_i~(i=1,2,3)$ can be directly verified by putting them in Eq.(1) through Maple. Here, we also discuss the dynamics properties of these nonlinear waves in six cases.

(\textrm{i}) When $\alpha=\beta=0$. Choosing $d_1=1,d_2=-2,d_3=2,\epsilon=\dfrac{1}{100}$, we can get the following concrete expressions. In this case, the three components $q_1, q_2$ and $q_3$ are also proportionable to each other, thus they are the second-order RW. When $m[1]=n[1]=0$, they are the fundamental second-order RW in Fig.13; conversely, $m_1 \neq0, n_1\neq0$, they are the second-order RW of triangular pattern in Fig.14.
\begin{eqnarray}
&& q_1[2]=3e^{9it}\dfrac{15000it-69687t^2+8100xt-7500x^2+625}{r_1}-2e^{9it}\dfrac{ip_1+p_2}{r_1r_2},\\
&&q_2[2]=-6e^{9it}\dfrac{15000it-69687t^2+8100xt-7500x^2+625}{r_1}+4e^{9it}\dfrac{ip_1+p_2}{r_1r_2},\\
&&q_3[2]=6e^{9it}\dfrac{15000it-69687t^2+8100xt-7500x^2+625}{r_1}-4e^{9it}\dfrac{ip_1+p_2}{r_1r_2},
\end{eqnarray}
where
\begin{eqnarray*}
&& r_1=209061t^2-24300tx+22500x^2+625,\\
&&r_2=1015258328477109t^6-354022663940100t^5x+368948239537500t^4x^2-77797057500000t^3x^3\\
&&\quad +39707718750000t^2x^4-4100625000000tx^5+1265625000000x^6+105468750000x^4+\\
&&\quad75937500000x^3t-6002859375000x^2t^2-1851022125000xt^3+79615779001875 t^4-\\
&&\quad 2636718750000m_1x^3+(4271484375000m_1-23730468750000n_1)x^2t+(25628906250000n_1+\\
&&\quad 68884804687500m_1)xt^2+(64271601562500n_1-38028171093750m_1)t^3+26367187500x^2-\\
&&\quad53789062500xt+909699609375t^2+219726562500m_1x+(1977539062500n_1-\\
&&\quad435058593750m_1)t+1373291015625m_1^2+1373291015625n_1^2+244140625,\\
&&p_1=-31862039754609000000t^6x+33205341558375000000t^5x^2-7001735175000000000t^4x^3\\
&&\quad+3573694687500000000t^3x^4-369056250000000000t^2x^5+113906250000000000tx^6+\\
&&\quad 91373249562939810000t^7-9492187500000000x^4t+6834375000000000x^3t^2-\\
&&\quad 165323531250000000x^2t^3-190506836250000000xt^4+245741455378125000t^5+\\
&&\quad(118652343750000000m_1+64072265625000000n_1)x^3t-(192216796875000000m_1+
\end{eqnarray*}
\begin{eqnarray*}
&&585834082031250000n_1)t^2x^2+(595333863281250000n_1+3307410351562500000m_1)xt^3-\\
&&(1748634644531250000m_1-3846298727285156250n_1)t^4-791015625000000tx^2-\\
&&-284765625000000xt^2+12659730468750000t^3-1647949218750000n_1x^2+\\
&&(1779785156250000n_1+9887695312500000m_1)xt+(44014086914062500n_1-\\
&&12458496093750000m_1)t^2+(30899047851562500m_1^2 t+30899047851562500n_1^2)t-\\
&&21972656250000t-22888183593750n_1,\\
&&p_2=-28476562500000000x^8+123018750000000000x^7t-1257661687500000000 x^6 t^2+\\
&&\quad 3572612122500000000x^5t^3-33195282797421000000t^5x^3+18493139276437500000t^4x^4\\
&&\quad-108578751030428670000x^2t^6+98683109527974994800xt^7-212250921409752884649t^8+\\
&&\quad 3164062500000000x^6-17085937500000000tx^5+459440859375000000t^2x^4-\\
&&\quad 829515431250000000x^3t^3+4407181779093750000x^2t^4-2656332241017750000xt^5+\\
&&\quad2548145144013862500t^6+59326171875000000x^5m_1+(533935546875000000n_1-\\
&&\quad160180664062500000m_1)tx^4-(894875976562500000m_1+1153300781250000000n_1)t^2x^3+\\
&&\quad(1636533808593750000m_1+4137786914062500000n_1)t^3x^2-(15325210710351562500m_1+\\
&&\quad3796204851562500000n_1)t^4x+(7950207477030468750 m_1-13436685294257812500 n_1)t^5+\\
&&\quad395507812500000x^4-1233984375000000x^3t+8016152343750000 x^2 t^2-18078630468750000 x t^3\\
&&\quad+121360886107031250 t^4+3295898437500000 m_1x^3+(1779785156250000m_1\\
&&\quad-29663085937500000n_1)tx^2+(84183837890625000m_1+32036132812500000 n_1)xt^2-\\
&&\quad(17575022460937500m_1-97639013671875000n_1)t^3-(30899047851562500m_1^2+\\
&&\quad30899047851562500n_1^2+21972656250000)x^2+(33370971679687500m_1^2 -71191406250000\\
&&\quad+33370971679687500n_1^2)xt-(287101593017578125m_1^2+287101593017578125n_1^2-\\
&&\quad 1032196289062500)t^2+411987304687500 m_1x-(420227050781250m_1-\\
&&\quad 2059936523437500n_1)t+858306884765625m_1^2+858306884765625n_1^2-152587890625.
\end{eqnarray*}


(\textrm{ii}) When one of these two parameters $\alpha$ and $\beta$ is zero and $q_i\neq0~(i=1,2,3)$, for convenience, we consider $\alpha=0, \beta\neq0$. Here, we can see that this combination in Fig.15, $q_1$ and $q_3$ are all the fundamental second-order RW and two-breather, and $q_2$ is the second-order RW and two-dark soliton. Increasing the absolute values of $\beta$, it reveals that the fundamental RW merges with breathers (dark solitons) distinctively from Fig.16. Given that these two constants $m_1$ and $n_1$ are all not zero, the phenomenon that, the seconder-order RW in Fig.15 splits into three first-order RW and these three humps form a triangle, is displayed in Fig.17.

(\textrm{iii}) When $\alpha=0,\beta\neq0$ and $d_1\neq0, d_2\neq0,d_3=0$, we can find that this combination, $q_1$ and $q_2$ are all the fundamental second-order RW and two-dark soliton, $q_3$ is the fundamental second-order RW and two-bright soliton in Fig.18. The character that the fundamental RW integrates with solions distinctively is shown in Fig.19. We discover that the fundamental second-order RW splits into three first-order RW from Fig.20.

.

(\textrm{iv}) When $\alpha\neq0, \beta\neq0$ and $d_1\neq0, d_2\neq0, d_3=0$, we can observe that $q_1$ and $q_2$ are all the fundamental second-order RW and two-breather, $q_3$ is the fundamental RW and two-bright soliton in Fig.21. Here,the rogue wave in $q_3$  and the triangular pattern in Fig.23(c) are not discovered for the same reason as the first-order rogue wave. We can find an interesting phenomenon that the plane which the RW and breather emerge is high, and the background plane where these nonlinear waves disappear is low. It is the same as the case in (\textrm{iii}), and the nonlinear waves merge with each other distinctly in Fig.22. From Fig.23, we can also see that the second-order RW of triangular pattern.

(\textrm{v}) When $\alpha \neq0, \beta \neq0$, and $d_1\neq0, d_2=d_3=0$, we arrive at the interactional solutions between two-dark (two-bright) soliton and seconder-order RW. In Fig.24, it reveals that $q_1$ is the fundamental second-order RW and two-dark soliton, $q_2$ and $q_3$ are all the fundamental second-order RW and two-bright soliton. Here, the rogue waves in $q_2$ and $q_3$ are not obvious to appear for the same reason as the first-order rogue wave. By increasing the absolute  values of $\alpha$ and $\beta$, we see these interactional solutions merge with each other from Fig.25. Analogously, in Fig.26(a), the second-order RW of triangular pattern is easily observed, in Fig.26(b) and (c), the triangular pattern is not easy to find because of their small amplitude. It is shown that the interactional process of Fig.26 is also elastic, and is similar to the first-order localized waves, thus the amplitudes and velocities of these two dark and bright solitons remain unchanged after collision in Fig.27.

(\textrm{vi}) When $\alpha \neq0, \beta \neq 0$, the $d_i~(i=1,2,3)$ are all not zero. These three components $q_1, q_2,$ and $q_3$ are all the fundamental second-order RW and two-breather in Fig.28. In the same way, increasing the absolute values of $\alpha$ and $\beta$, we can see the second-order RW merge with two-breather distinctively in $q_1$, $q_2$ and $q_3$ from Fig.29. Besides, in Fig.30, the fundamental second-order RW splits into three first-order RW.

In the first-order localized waves, we get the concrete expressions of these interactional solutions and give the classifications in three different cases. Instead of considering various arrangements of the three potential functions $q_1,q_2$ and $q_3$, the conclusion that there are six types first-order localized waves, which are obtained using our method in this paper, can be drawn. However, the expressions of the second-order localized waves are greatly tedious and complicated, we can not give these expressions in the general form. Then, the classifications as the first-order ones are also not presented, and we give the six types solutions which are similar with the first-order case. Whether the second-order localized waves own more types or not, we can not draw a firm conclusion now.

\section{Conclusions}
We arrive at some interesting and appealing localized nonlinear waves in the three-component coupled Hirota equation by the generalized Darboux transformation. By choosing a periodic seed solution of Eq.(1), we can get a peculiar vector solution of Lax pair of Eqs.(2) and (3). With a fixed spectral parameter and this special vector solution, we implement  the Taylor series expansion of Eq.(28) at $f=0$, and construct the generalized Darboux transformation of this coupled system Eq.(1). So, we can get its multi-parametric and semi-rational solutions, and some free parameters play the important role in controlling the dynamic properties of these localized nonlinear waves, such as $\alpha, \beta, d_i, s_i~(i=1,2,3)$.

We mainly discuss the dynamics of these interactional solutions in six diverse cases. (\textrm{i}) $\alpha=\beta=0$, the first- and the second-order rogue wave are exhibited in these three components $q_1, q_2$ and $q_3$. Then, by choosing $s_1\neq0$, the fundamental second-order rogue wave splits into three first-order one, and they form a triangle. (\textrm{ii}) $\alpha=0, \beta\neq0$, we present the first- and the second-order breather-dark-rogue wave solutions. The rogue wave coexists with breather in $q_1$ and $q_3$, and $q_2$ is the rogue wave and  dark soliton. Increasing the absolute values of $\alpha$ and $\beta$ and choosing $s_1\neq0$, these phenomena, the nonlinear waves merge with each other and the second-order rogue wave splits into triangular pattern, are all presented respectively. (\textrm{iii}) $\alpha=0,\beta\neq0, d_1\neq0, d_2\neq0, d_3=0$, the first- and the second-order dark-bright-rogue wave solutions are exhibited. If we also increase the absolute values of $\alpha$ and $\beta$, we can see that the nonlinear waves integrate with each other. Choosing $s_1\neq0$, the fundamental second-order rogue wave splits into three first-order one. (\textrm{iv}) $\alpha\neq0, \beta\neq0, d_1\neq0,d_2\neq0, d_3=0$, the first- and the second-order breather-dark-rogue wave solutions are reached. Changing the $\alpha,\beta$ and $s_1$, these similar cases in  (\textrm{ii}) are also presented here. (\textrm{v}) $\alpha\neq0, \beta\neq0, d_1\neq0,d_2=d_3=0$, the first- and the second-order dark-bright-rogue wave solutions are also derived. Similarly, the roles of $\alpha, \beta$ and $s_1$ in determining dynamics can be also exhibited. (\textrm{vi}) $\alpha\neq0,\beta\neq0,d_1\neq0,d_2\neq0,d_3\neq0$, we can construct that the first- and the second-order breather-rogue wave solutions. Here, these three components are all the rogue wave interplays with breather. Then, the effects which caused by $\alpha, \beta$ and $s_1$ are reached similarly.

In this paper, we generalize Baronio's results\cite{26}, thus reach the higher-order localized waves of the three-component coupled system by the generalized Darboux transformation. Using Darboux transformation, one gets the rogue wave and dark-breather-rogue wave in two-component coupled Hirota equation\cite{30}. In \cite{31}, the higher-order localized waves are been figured out. Here, we extend the two-component system to three-component one, then the Lax pair become $4\times4$ matrix, and the more free parameters may exist in vector solution of this kind of Lax pair in Eq.(28). Combining this vector solutions with multi parameters and the generalized Darboux transformation, we arrive at some new combination of these interactional solutions, such as Type 2, 3, 4 and 5 in Table 3 in the first-order localized waves and similar cases in the second-order localized waves. However these solutions are not possible to be obtained in the two-component one\cite{31}. Through considering both two-component(even) and three-component (odd) coupled Hirota equation, we may well understand  the localized waves of the multi-component coupled Hirota equation\cite{32}. Furthermore, we expect these localized waves in this paper will be verified in physical experiments in the future.

\section*{Acknowledgment}
We would like to express our sincere thanks to our discussion group for their discussions and suggestions. The project is supported by the Global Change Research
Program of China (No.2015CB953
904), National Natural Science Foundation of China (No. 11275072, 11435005 and 11675054) and Shanghai Collaborative Innovation Center of
Trustworthy Software for Internet of Things (No. ZF1213).

\end{document}